\documentclass[twocolumn,showpacs,preprintnumbers,amsmath,amssymb]{revtex4}
\usepackage{dcolumn}
\usepackage{bm}
\usepackage[dvipdf]{graphicx}
\DeclareGraphicsExtensions{.jpg,.pdf,.mps,.png,.eps,.ps,.EPS}
\begin{document}
\newcommand{\avg}[1]{\langle{#1}\rangle}
\newcommand{\Avg}[1]{\left\langle{#1}\right\rangle}
\def\be{\begin{equation}}
\def\ee{\end{equation}}
\def\bc{\begin{center}} 
\def\ec{\end{center}}
\def\bea{\begin{eqnarray}}
\def\eea{\end{eqnarray}}

\title{Effect of degree correlations on the loop structure of scale-free networks}
\author{Ginestra Bianconi and Matteo Marsili}
\affiliation{The Abdus Salam International Center for Theoretical Physics, Strada Costiera 11, 34014 Trieste, Italy } 

\begin{abstract}
 In this paper we study the impact  of degree correlations  in the subgraph statistics of scale-free networks. In particular we  consider loops, simple cases of network subgraphs which encode the redundancy of the paths passing through every two nodes of the network.
We   provide an  understanding of the scaling of the clustering coefficient in modular networks in terms of the maximal eigenvector of the average adjacency matrix of the ensemble. Furthermore we  show that correlations affect in a relevant way the average number of Hamiltonian paths in a three-core of real world networks.
We  prove our results in the  two-vertex correlated hidden variable ensemble and we check the results with exact counting of small loops in real graphs.
\end{abstract}

\maketitle 
\section{INTRODUCTION}
The dynamics and the function of many complex systems strongly affect their network structure \cite{RMP,Doro,Newman,Vespignani}.
In fact both large-scale properties (like the scale-free degree distribution \cite{BA}) and local properties (like recurrence of small motifs \cite{Milo,Vazquez}) must  be selected for widespread robustness requirements and specific preferential uses in real graphs.
A large number of  different  networks \cite{RMP,Doro,Newman}, from the Internet to the protein interaction networks in a cell, share a  scale-free degree distribution $P(k)\sim k^{-\gamma}$ with $\gamma<3$  and a high clustering coefficient respect to  random Erd\"os-Renyi graphs \cite{Janson}.  
The scale-free degree distribution  of a network   affects the statistics of subgraphs present in it showing that large-scale properties and local properties of scale-free networks are strongly related to each other.
Special examples of subgraphs in networks are loops \cite{loop1,large}, paths that pass through each node in the loop only once.  
In random scale-free networks  there are many small size loops compared to random graphs and there can be a lack of Hamilton cycles (loops of length $L=N$) due to the fact that most of the large paths need to pass through hubs \cite{large}.
Along with other properties, many real scale-free networks also have degree correlations \cite{Berg}. Degree correlations in real networks indicate that links are not randomly wired and that the  probability that two nodes of degree $k_i$ and $k_j$ are linked deviates from the expected value $r_{i,j}=k_i k_j/( \langle{k}\rangle N)$. Consequently, correlated networks have at least  one of the three following  features: {\it i)} a $k$ dependent average connectivity $k_{nn}(k)$ of the first neighbors of a node with degree $k$  \cite{Ass1,Ass2}; {\it ii)} a non trivial dependence on the connectivity of the clustering coefficient $C=C(k)$ of nodes of degree $k$  \cite{ravasz}; {\it iii)} a cutoff that is larger than the structural cutoff $K\sim \sqrt{\langle k\rangle N}$.
In particular many real scale-free networks  show a power-law dependence on $k$ both for $ k_{nn}(k)$ and for $C(k)$, i.e. $k_{nn}(k)\sim k^{\alpha}$ and $C(k)\sim k^{-\delta}$. 
Correlations do affect the subgraph statistics as shown in the Internet  \cite{Internet} and in calculations based on the scaling of the clustering coefficient \cite{Vazquez,Vazquez2}.
Every network can be represented in terms of its adjacency matrix $((a))$ of elements $a_{i,j}=1,0$ depending if there is a link between node $i$ and node $j$. 
From a formal point of view an ensemble of networks is given when a probability ${\cal P}(a)$ is assigned to each adjacency matrix $((a))$ of $N\times N$ elements.
In an uncorrelated and undirected network ensemble with given  degree sequence $\{k_i\}$  all the links are independent. Consequently all the matrix elements $a_{i,j}$ with $i<j$ of are independent and their average value in the ensemble  can be written as $\langle a_{i,j}\rangle=r_{i,j}=\frac{k_i k_j}{\langle k \rangle  N}$.
A two-vertex correlated network is a network in which still the matrix elements   $a_{i,j}$ with $i<j$  are independent but $\langle a_{i,j}\rangle =r_{i,j}\neq \frac{k_i k_j}{\langle k \rangle N}$.
Networks with higher order correlations  instead would have non independent matrix elements which will favor some specific motifs in the network.  
In this paper we are going to provide an analytic calculation of the number of loops in two-vertex correlated scale-free networks. 
In the light of our results we are able  to interpret the  scaling of the clustering coefficient $C(k)$ in terms of  the scaling of  the maximal eigenvector (the eigenvector associated with the maximal eigenvalue) of the average adjacency matrix of the network ensemble. 
Moreover we show that the maximal eigenvalue and the corresponding eigenvector  not only determine the number of  triangles in the two-vertex correlated network ensemble, but they  also fix the number of small loops of length $3\ll L\ll N$.
Finally  we are able to give a sufficient condition  for the absence of Hamilton cycles in two-vertex correlated networks. This allows us to study a set of real graphs (the Internet at the Autonomous System -AS- Level  and protein-protein interaction networks-DIP) \cite{note2} and show that assuming they are specific instances of  two-vertex correlated network ensembles, one can exclude the presence of  Hamiltonian cycles in  the three-core of these graphs. Our findings are  in agreement for the Internet with  what was found in Ref. \cite{gui} where a Belief-Propagation algorithm was applied to the measurement of the number of loops in real graphs. The absence of  Hamiltonian cycles in a three-core of a network is an unexpected result since  regular random graphs with connectivity $c \ge 3$ are Hamiltonian \cite{Janson,Monasson}. 
We note here that  the average number $\avg{{\cal N}_L}$ of loops of size $L$ in a two-vertex correlated network can possibly be dominated by a  very large number of loops occurring in very rare networks \cite{gui}. Nevertheless, preliminary results  indicate that  in uncorrelated scale-free networks with $\gamma<3$   the ratio $\frac{\avg{{\cal N}_L^2}}{\avg{{\cal N}_L}^2}$ is bounded at least for  small loops and for Hamiltonian cycles. We expect that  similar arguments could  also be extended  to scale-free  correlated networks.

The paper is organized as follows: in Sec. II we give an intuition of the results found for small loops in two-vertex correlated network ensemble by considering the problem of exact counting of loops in generic networks;
in Sec. III we introduce the hidden variable ensemble and  we calculate the number of small loops and Hamiltonian cycles in
two vertices correlated hidden variable ensembles;
in Sec. IV we compare the results with real networks; and finally  we give the conclusions in Sec.V.

\section{Counting  small loops in real networks}
In this section we would like to provide some intuitive arguments to show that in scale-free networks  the maximal eigenvalue of the adjacency matrix  and the corresponding eigenvector are  responsible
for the number of small loops present in it.
The adjacency matrix $((a))$ of a simple network of size $N$ is the $N\times N$ matrix of elements $a_{i,j}=1,0$ indicating the existence ($a_{i,j}=1$) or not ($a_{i,j}=0$ ) of a   link between node $i$ and node $j$.  
The total number of closed paths  of length $L$ 
passing though a node $i$ is given by the matrix element $(a^L)_{i,i}$.
The loops ${\cal N}_L^{(i)}$  of size $L$  passing through a node $i$ are given
by
\be
{\cal N}_L^{(i)}=(a^L)_{i,i}-(\mbox{corrections})
\ee 
where these corrections account for closed paths which intersect themselves at least once and which must be subtracted from the term $(a^L)_{i,i}$ in order to consider only loops.
  If by $\lambda_n$ we indicate  the eigenvalues and by ${\bf u}^{n}$  the eigenvectors of the adjacency matrix $((a))$ we  find \cite{loop1}
\be
{\cal N}_L^{(i)}\sim \sum_n \lambda^L_n \left[u^{(n)}_i u^{(n)}_i- {\cal O}(u^{(n)4}_i)\right]
\label{exn}
\ee 
 For  small $L$, the  correction terms can be neglected if the spectrum of the graphs $\{\lambda\}$ contains one large eigenvalue $\lambda_0=\Lambda_0$ and if the  associated normalized eigenvectors satisfy $0< u^{(n)}_i \ll 1 $,$\forall i$, as  is the case in  most scale-free networks. If these conditions are satisfied the sum over $n$ in $(\ref{exn})$ is dominated by  the term $n=0$ and consequently the number of loops of length $L$ passing through the node $i$ is given by
\be
{\cal N}_L^{(i)}\sim  \Lambda^L_0 u^{0}_i u^{0}_i ,
\label{nliai}
\ee
while the total number of loops of size $L$ is given by
\be
{\cal N}_L=\frac{1}{2L}\sum_i{\cal N}_L^{(i)}\sim \frac{\Lambda_0^L}{2L},
\label{nlia}
\ee
where the factor $2L$ accounts for the multiplicity of nodes a single loop pass through and the two possible directions of each loop. 
Thus we found by intuitive arguments that  the total number of small loops  of size $L$ of scale-free networks will scale like $\Lambda^L_0$ while the number of small loops passing through a node is proportional to the square of the maximal eigenvector associated with $\Lambda_0$.
These  arguments apply for the exact counting of  small loops in real networks.
In a random graph ensemble the adjacency matrix is a random variable which has average values of the elements $\avg{a_{i,j}}=r_{i,j}$ and we need to evaluate the average number of loops   $\Avg{{\cal N}_L}$ instead of   ${\cal N}_L$
The results we will prove in the following sections are an extension of the expressions $(\ref{nlia})$ and $(\ref{nliai})$ to two-vertex correlated hidden variable network ensemble.

\section{ Average number of loops in correlated hidden variable ensemble } 
To model a general two-vertex correlated network in the following we will consider networks that are generated within  the hidden variable model  \cite{HV1,HV2}.
The prescription of Ref. \cite{HV1} to generate a class of  scale-free networks with exponent $\gamma$ is the following: {\em 1)} assign to each node $i$ of the graph a hidden continuous variable $q_i$ distributed according to a $\rho(q)$ distribution. Then {\em 2)} each pair of nodes with hidden variables $q,q'$ are linked with probability $r(q,q')$.
When the hidden variable distribution is scale-free $\rho(q)=\rho_0 q^{-\gamma}$ for $q\in [m,Q]$  and $r(q,q')=qq'/(\avg{q}N)$,  we obtain a random uncorrelated scale-free network.
In this specific case  a structural cutoff   is needed to keep the linking probability smaller than one, i.e. $Q^2/(\avg{q}N)<1$. This cutoff scales differently with the system size $N$ depending on the value of $\gamma$: $Q\sim N^{1/(\gamma-1)}$ for $\gamma>3$, $Q\sim  N^{1/2}$ for $\gamma\in (2,3)$ and $Q\sim N^{1/\gamma}$ for $\gamma\in (1,2)$. 
On the contrary, to generate a correlated scale-free network with natural cutoff $N^{1/(\gamma-1)}$ and  $\gamma>2$ in the literature different ansatz have been proposed  \cite{HV1,HV2}.
 In order to   present  general results on the average number of loops in the hidden variable ensemble for any type of linking probabilities  $r(q,q')$ we consider  an   ordered set of distinct nodes $\{i_1,\ldots,i_{n},\ldots,i_L\}$.  With each such kind of set it is  possible to associate a loop in the network in which subsequent nodes are linked with each other. For each choice of the nodes $\{i_1,\ldots,i_L\}$, with hidden variables $\{q_{i_1},\dots,q_{i_L}\}$ the
probability that they are connected in a loop is
\be
r(q_{i_1},q_{i_2})r(q_{i_2},q_{i_3})\cdots
r(q_{i_L},q_{i_1})=\prod_n r(q_{i_n},q_{i_{mod(n+1,L)}})
\label{rql}
\ee
and for each loop of the network there are $2L$ ordered sets $\{i_1,\ldots,i_L\}$ which describe it corresponding to cyclic permutations of the indices and to their order inversion.
The average number of loops of size $L $ in the graph is given by the number of ways we can choose an ordered set of $L$ nodes  $\{i_1,\ldots,i_L\}$ multiplied by the probability that these nodes are connected in all distinguishable orderings and divided by $2L$.
 In order to proceed with the calculation, we lump together nodes with hidden variable $q_i\in[q,q+\Delta q)$, where $\Delta q $ is a small interval of $q$. In each interval of $q$ there are $N_q\simeq NP(q)\Delta q$ nodes of the network.  For each choice of the $L$ nodes, let $n_q$ with $\sum_q n_q=L$ be the number of nodes in the loop with $q_{i_n}\in [q,q+\Delta q)$. The ways we can choose them  within the $N_q$ nodes of the network, is given by the binomial $N_q!/[n_q! (N_q-n_q)!]$.  Moreover let   ${n_{q,q'}}$ indicate the nodes of a hidden variable $q$ of the loop linked with a subsequent  node of hidden variable $q'$ in the fixed direction of the loop. We note that the way to choose $\{n_{q,q'}\}$ is given by the multinomial $n_q!/\prod_{q'} n_{q,q'}!$ and that the partition $\{n_{q,q'}\}$ must satisfy the  conditions $\sum_{q'} n_{q,q'}=n_q$ and $\sum_q n_{q,q'}=n_{q'}$.
Finally the number of ways in which one can permute the $L$ nodes keeping ${n_{q,q'}}$ constant  is given by $\prod_q n_q!$.
Considering all this and that the probability  Eq. (\ref{rql}) that the selected nodes are connected in the chosen order  can be written as $\Pi_{q,q'} r(q,q')^{n_{q,q'}}$, we get  the following expression for the average number of loops $\Avg{{\cal N}_L}$ of size $L$, 
\begin{widetext}
\be
\Avg{{\cal N}_L}=\frac{1}{2 L} \sum_{\{n_q\}}'  \prod_q \frac{N_q!}{n_q! (N_q-n_q)!} \prod_q n_q! \sum_{\{n_{q,q'}\}}' \frac{n_q!}{\prod_{q'} n_{q,q'}!} \prod_{q,q'}{r(q,q')}^{n_{q,q'}}
\label{uno.eq}
\ee
\end{widetext}
where the sums  $\sum_{\{n_q\}}',\sum_{\{n_{q,q'}\}}'$ are extended over all $\{n_q\}$ and $\{n_{q,q'}\}$ such that $\sum_q n_q=L$, $\sum_{q'} n_{q,q'}=n_q$ and $\sum_q n_{q,q'}=n_{q'}$ and the factor $2L$ accounts for the multiplicity in which we count each loop. Introducing the constraints $\sum_q n_q=L$ and $\sum_q n_{q,q'}=n_{q'}$ by explicit   delta functions, using their integral representation we find
\begin{widetext}
\bea
\Avg{{\cal N}_L}&=&\nonumber\frac{1}{2 L} \int_{-\infty}^\infty dx\ \sum_{\{n_q\}}  e^{Lx} \prod_q \frac{N_q!}{n_q! (N_q-n_q)!} \prod_q n_q! e^{-x n_q}  \int_{-\infty}^\infty {\cal D}x_q  \ \prod_q e^{ n_q x_q} \sum_{\{n_{q,q'}\}} \frac{n_q!}{\prod_{q'} n_{q,q'}!} \prod_{q,q'}{r(q,q')}^{n_{q,q'}} e^{- x_{q'}n_{q,q'}}.
\eea
\end{widetext}
where the ${\cal D}x_q$ indicates $\prod_q d x_q$, and the sum over $\{n_{q,q'}\}$ is performed over all $\{n_{q,q'}\}$ such that $\sum_{q'} n_{q,q'}=n_q$.
Consequently, performing the multinomial summations over  $\{n_{q,q'}\}$ we get the following expressions:
\begin{widetext}
\bea
\Avg{{\cal N}_L} &=&\frac{1}{2 L}
\int_{-\infty}^\infty dx\   e^{Lx} \sum_{\{n_q\}} \prod_q \frac{N_q!}{n_q! (N_q-n_q)!} e^{-x n_q} {n_q!} \int_{-\infty}^\infty {\cal D}x_q \prod_q e^{ n_q x_q} {\left(\sum_{q'} r(q,q')e^{- x_{q'}}\right)}^{n_q}\nonumber \\
&=& \frac{1}{2 L}
\int_{-\infty}^\infty dx\   e^{Lx} \sum_{\{n_q\}} \prod_q \frac{N_q!}{n_q! (N_q-n_q)!} e^{-x n_q} {n_q!} \int_{-\infty}^\infty {\cal D}x_q\  e^{Q g(\{x_q\})} 
\label{exact.eq}
\eea
\end{widetext}
with
\be
g(\{x_q\})=\frac{1}{Q}\sum_q n_q \left[x_q+\ln\left( \sum_{q'} r(q,q')e^{- x_{q'}} \right)\right]
\ee
Notice that in Eq. (\ref{exact.eq}) one can safely take the limit $\Delta q\rightarrow 0$ and that the average over the $P(q)$ distribution is taken assuming that we focus on the limit $N\to\infty$. In what follows, we will evaluate Eq. (\ref{exact.eq}) in different ranges of $L$ in the limit $N\to\infty$.
Assuming $L\gg 1$ we  evaluate the integral over the variables $\{x_q\}$ by the saddle point equation finding
\be
n_q=e^{-x_q}\sum_{q'} n_{q'} \frac{r(q',q)}{\sum_{q^{''}} r(q',q'')e^{-x_{q''}}}.
\ee
If we indicate by $S_{q'}$ the sum $S_{q'}=\sum_{\bar{q}} r(q',{\bar{q}}) e^{-x_{\bar{q}}}$, we can cast the solution in the following form,
\be
e^{-x_q}=n_q\frac{1}{\sum_{q'} n_{q'}{r(q,q')}/{S_q'}}.
\label{x_q}
\ee
This provides the self-consistent equation for $\{S_q\}$
\be
S_q=\sum_{q'} n_{q'}\frac{r(q,q')}{\sum_{q''}n_{q''} {r(q',q'')}/{S_{q''}}}
\label{S_q.sceq}
\ee
It is easy to  check that  $\{S_q\}$ satisfying   the equation
\be
S_q=\sum_{q'} n_{q'} {r(q,q')}/{S_{q'}}
\ee
is a  solution of the Eq. $(\ref{S_q.sceq})$.
Inserting a delta function $\delta \left(S_q-\sum_{q'}n_{q'}{r(q,q')}/{S_{q'}}\right)$ and assuming that the Jacobian of this transformation is $1$, i.e. assuming 
\be S_q^2 \gg r(q,q')
\label{condition}
\ee and using the Stirling approximation for the factorial $n_q$, the integrals over $x_q$ calculated at the saddle point take the values $S_q^{2n_q}e^{-n_q\ln(n_q)+n_q}$ and  the average number of loops of size $L$ can be expressed as the following:
\begin{widetext}
\bea
\Avg{{\cal N}_L}&=&\int_{-\infty}^\infty dx\   e^{L(x-1)}  \int {\cal D}S_q \ \int {\cal D}w_q \  \sum_{n_q}' \prod_q \frac{N_q!}{n_q! (N_q-n_q)!}{\left(e^{-x}{S_q}^2\right)}^{n_q} \exp\left[{\sum_q w_q \left(S_q-\sum_{q'}n_{q'}\frac{r(q,q')}{S_{q'}}\right)}\right]. 
\label{avgsq0}
\eea
\end{widetext}
Finally, performing the summation over $\{n_q\}$ we get 
\begin{widetext}
\bea
\Avg{{\cal N}_L}&=&\frac{1}{2 L}\int_{-\infty}^\infty dx\   e^{L(x-1)} \prod_q \int {\cal D}S_q \ \int {\cal D}w_q \   \exp\left\{{N\left\langle \ln(1+e^{-x}{S_q}^2\exp\left[{-N \sum_{q'} w_{q'} r(q,q') /S_{q}}\right]\right\rangle+ N \sum_q w_q S_q}\right\}.\nonumber
\label{avgsq}
\eea
\end{widetext}
where $\avg{}_q$ indicates the average over the distribution of the hidden variables $N_q$.
In the limit $N> L\gg 1$ we evaluate the  saddle point equations, finding  
\begin{widetext}
\bea
S_{q}&=&N \left\langle{ \frac{r(q,q')S_{q'} e^{-x}\exp\left(-\sum_{q^{''}}r(q',q'')w_{q''}/S_{q'}\right)}{1+S^2_{q'}e^{-x}\exp\left({-\sum_{q^{''}}r(q',q'')w_{q''}/S_{q'}}\right)}}\right \rangle_{q'} \nonumber \\
w_{q}&=&- P(q) {\frac{(2S_{q}+\sum_{q'}r(q,q')w_{q'})e^{-x}\exp\left({-\sum_{q^{'}}r(q,q')w_{q'}/S_q}\right)}{1+S^2_{q}e^{-x}\exp\left({-\sum_{q^{'}}r(q,q')w_{q'}}/S_q\right)}} \nonumber \\
\ell&=&\left\langle {\frac{S^2_qe^{-x}\exp\left({-\sum_{q'}r(q,{q'})w_{q'}/S_q}\right)}{1+S^2_qe^{-x}\exp\left({-\sum_{q'}r(q,{q'})w_{q'}/S_q}\right)}} \right\rangle_{q}. 
\label{sp.eq}
\eea
\end{widetext}
with $\ell=L/N$.
In order to solve these saddle point equations we make the ansatz 
\be
N \sum_{q'}r(q,q')w_{q'}=\nu S_q.
\ee
With this assumption we can  rewrite the saddle point equations $(\ref{sp.eq})$ as
\bea
S_{q} &= &N \left\langle{ \frac{r(q,q')S_{q'} e^{-x-\nu}}{1+S^2_{q'}e^{-x-\nu}}}\right\rangle_{q'}, \nonumber \\
w_{q} &= &-(2+\nu) P(q) {\frac{S_{q}e^{-x-\nu}}{1+S^2_{q}e^{-x-\nu}}}, \nonumber \\
\ell &= &\left\langle {\frac{S^2_qe^{-x-\nu} }{1+S^2_qe^{-x-\nu}} }\right\rangle_{q} ,
\label{sp2.eq}
\eea
which can be solved and define the value of $\nu$, $\nu=-1$.
\begin{figure}
\includegraphics[width=75mm, height=65mm]{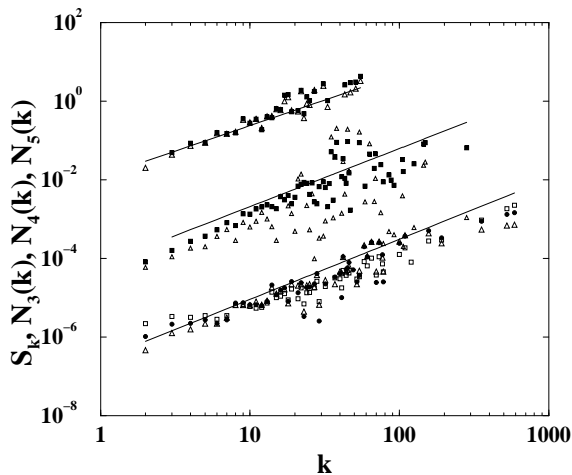}
\caption{\label{ck.fig} Normalized number of triangles (empty triangles), quadrilaterals (filled squares) and pentagons (filled circles), passing through nodes of connectivity $k$. Data are shown for  the Internet at the Autonomous System Level in November 1997 (bottom), in the s. cerevisiae  protein interaction database (center) and in the h. pylori protein interaction (top) \cite{note2}. The solid lines indicate the predictions  ${\cal N}_L(k)\propto S^2_k$ where $S_k$ is the maximal eigenvector of the correlation  matrix $ N_{k'} r(k,k')$. The data are shifted to improve the readability of the graph.
} 
\end{figure}
\subsection{ The uncorrelated case } In the uncorrelated case, when $r(q,q')=\frac{q q'}{\langle{q}\rangle N}$ we found $S_q=q\sqrt{\frac{\ell}{\langle{q} \rangle} }$ which satisfies hypothesis $(\ref{condition})$.
 The  results found in this limit are the same as the ones found in  \cite{large}.

\subsection{Small loops}
The limit of small loop size is the limit  $x\gg 1$.
In this limit the saddle point equations $(\ref{sp2.eq})$ reduce to 
\bea
S_q&=&\sum_{q'} N_{q'} r(q,q') S_{q'} e^{-x+1}\nonumber \\
w_q&=&-P(q)r(q,q') S_q e^{-x+1} \nonumber \\
\ell&=&{\langle S_q^2\rangle}_q e^{-x+1}. 
\eea
The first equation indicates that $S_q$ is the eigenvector of the average adjacency  matrix $ N_{q'} r(q,q')$ with eigenvalue $\Lambda=e^{x-1}$; the second equation defines the linear relation between $w_q$ and $S_q$, and the third equation fixes the normalization constant for the eigenvector $S_q$.
In this limit the average number of loops of size $L$ is given by
\be
{\cal N}_L \sim \frac{1}{2L}{\left( {\Lambda}\right)}^L
\ee
where $\Lambda$ is the maximal eigenvalue of the average adjacency  matrix $  N_{q'} r(q,q')$, with   the results  valid until
\be
\ell \ll \frac{\avg {S^4_q}}{\avg{ S^2_q}^2},
\ee
where $S_q$ is the eigenvector of matrix $N P(q')r(q,q')$ corresponding to the maximal eigenvalue $\Lambda\gg \max S_q^2$.
We observe that the vector $S_i=S_{q_i}$ with $i=1,\dots, N$ is the eigenvector of the matrix $r_{i,j}=r(q_i,q_j)$. In other words $\{S_i\}$ is the eigenvector of the average adjacency matrix of the networks in the ensemble $\avg{a_{i,j}}=r_{i,j}$. This result provides the extension of the arguments of Sec. I , Eq. $(\ref{nlia})$ to the two-vertex correlated network ensemble.
\begin{table}
\centering
\begin{tabular}{|l|l|c|c|}
\hline 
\multicolumn{2}{|l|}{Network } & $2\langle \ln(S_k) \rangle-$ & $2\langle \ln(S_k^R) \rangle -$ \\  
\multicolumn{2}{|l|}{ } & $\langle\ln(p)\rangle/N$ & $ \langle\ln(p^R)\rangle/N $ \\  \hline
AS & 11-97 & -4.73 & 2.98\\ \cline{2-4}
& 4-98 &-5.22& 3.06\\\cline{2-4}
 & 7-98 & -5.35 &3.03\\ \cline{2-4} 
& 10-98 & -5.56 & 3.01 \\ \cline{2-4}
& 1-99 & -5.74 & 3.07\\ \cline{2-4}
& 4-99 & -6.06 & 3.09 \\ \cline{2-4}
& 7-99& -6.28  &3.07 \\ \cline{2-4}
&10-99& -6.55 & 3.06\\ \cline{2-4}
& 1-00&-6.75 &3.07 \\ \cline{2-4} 
& 4-00&-7.20  &3.01 \\ \cline{2-4} 
& 7-00& -7.30& 3.03 \\ \cline{2-4} 
& 10-00& -7.46& 3.01\\ \cline{2-4}
& 1-01&-7.428 & 3.01 \\ \cline{2-4}
 & 3-01&-7.73 &3.00 \\ \cline{2-4}\hline
DIP 
&s. cerevisiase& -6.46 & 3.99 \\ \cline{2-4}
&h.pylori& -4.5 &3.8\\ \cline{2-4}
&c. elegans &-0.66 & 2.89 \\ \cline{2-4}\hline
\end{tabular} 
\caption{In the table we report the value of $ 2\langle \ln (S_q)\rangle-\langle\log(p)\rangle/N$ with $S_q$ satisfying Eq. $(\ref{hc2.eq})$ assuming as the maximum likelihood assumption that all the $q_i=k_i$ on the nodes of the three-core of the Internet graphs and on the graphs of protein interactions \cite{note2}. We compare the value of $2\langle \ln(S_q)\rangle-\langle\ln(p)\rangle/N$  calculated  with the   two-vertex correlation assumption on  real graphs or simply assuming  the minimal assumption  $r(q=k,q'=k')=1-e^{-k k'/\langle k \rangle N}$,i.e.  $2\langle \ln(S_q^R)\rangle-\langle\ln(p^R)\rangle/N$. We observe that real correlations are essential to predict the absence of  Hamiltonian cycles in these graphs. }
\label{data.table}
\end{table}

\subsection{ Small loops passing though a given node }
From expression (\ref{avgsq}) one can also derive the number of small loops passing through a given node. One can easily show that 
\be
{\cal N}_L(q)\sim \frac{1}{2L}S_q^2 \Lambda_{\{q\}}^{L-1}
\ee
where $S_q$ is the maximal eigenvector of the matrix $  N_{q'} r(q,q')$ normalized in such a way that $\langle S_q^2 \rangle=\ell \Lambda$.
This provides the extension of the arguments of section I Eq. $(\ref{nliai})$ to a two-vertex correlated network ensemble.
\subsection{ Hamiltonian cycles }
The Hamiltonian cycles of a graph are loops of size $L=N$. From Eq. (\ref{avgsq0})  we find that when $L=N$ the expected number of   Hamiltonian cycles goes to zero exponentially with  $N$  if 
\be
2 \langle{\ln(S_q)}\rangle<1
\label{hc1.eq}
\ee
with $S_q$ satisfying
\be
S_q=\sum_{q'}r(q,q'){N_{q'}}/{S_q'}.
\label{hc2.eq}
\ee
Consequently, in the thermodynamic limit, since
\be
P({\cal N}_L>0)\leq \Avg{{\cal N}_L},
\ee 
 $(\ref{hc1.eq})$ if a sufficient condition for excluding the presence of  Hamiltonian cycles in the network.

\section{ Comparison with real data }
To test our calculation on real graphs and forecast some results regarding the existence or not of  Hamiltonian cycles we have to assume that the real networks under study  are a particular instance of a  two-vertex correlated hidden variable network ensemble.
Since the average connectivity  $\bar k(q)$ of  a node  depends only on its hidden variable the minimal assumption one can make to fit real networks with the hidden variable model is that  the average degree is a one-to-one map to the hidden variable $q$.
In this assumption maximum likelihood considerations force us to assume that each real graph is a random realization of a two-vertex correlated networks with $q_i=k_i$  and $r(q=k,q=k')=\frac{N_{k,k'}}{\langle k \rangle N N_k N_{k'}}$ where $N_{k,k'}$ are the total number of links between nodes of degree $k$ and $k'$ and $N_k$ and $N_{k'}$ are the numbers of nodes with degree $k$ and $k'$. 

This results give a very interesting interpretation of the dependence of the clustering coefficient on the connectivity $k$, i.e. $
C(k)\sim\frac{1}{k(k-1)}   \Lambda_{\{k\}}^2 S_k^2 $ where $S_k$ is the  eigenvector associated with the maximal eigenvalue $\Lambda$ of the matrix $  N_{q'}r(q,q')$, in agreement with the intuitive arguments of Sec. I.
Moreover, one can predict if in the three-core of the considered graph there are no  Hamiltonian cycles by evaluating if the condition $(\ref{hc1.eq})$ is satisfied, i.e. if
\bea
2\langle{\ln(S_q)}-\frac{1}{N}\langle \ln(p)\rangle\rangle<1& \mbox{with} &
S_q=\sum_{q'}r(q,q'){N(q')}/{S_q'}.\nonumber
\label{hcrn}
\eea
where the $\ln(p)/N=\langle \ln[1-(1+q+q^2/2)e^{-q}]\rangle$ corrects for the probability that the network in the ensemble contains nodes of connectivity $k<3$ as described in \cite{large}.
In particular one can compare the value of $2\langle \ln{S_q}\rangle$ calculated by solving $(\ref{hcrn})$ with $r(q=k,q'=k')$   extracted from the data $[r(q=k,q'=k')=\frac{N_{k,k'}}{\langle k \rangle N  N_k N_{k'}}]$ with the value of $2\langle \ln{S_q}\rangle$ in  the  simplest example of a correlated ensemble, i.e. the static network ensemble \cite{Kahng} defined with  $r(q=k,q'=k')=1-\exp[-\frac{k k'}{\langle k\rangle N}]$.
We found as reported in Table $\ref{data.table}$ that the real degree correlations are such that the presence of Hamiltonian cycles in the three-core of the network is very unlikely.

\section{Conclusions}

In conclusion we have evaluated the number of loops of any size in two-vertex correlated networks. The results can be applied to real graphs,  finding very good agreement of the predicted scaling of the clustering coefficient $C(k)$ with the square of the maximal eigenvector $S_k$ of the  matrix $ N_{k'}r(k,k')$,i.e. $C(k)\sim S_k^2$. Moreover we can have a condition for predicting the absence of Hamiltonian cycles for the three-core of Internet and protein-protein interaction data. 
The results indicate that degree correlations strongly affect the loop frequency.
Further study  would consider how important are fluctuations of the number of loops around this average and would consider the frequency of other subgraphs
in correlated scale-free networks.   

\section{acknowlegments}
The work was supported by EVERGROW, integrated project No. 1935 in the
complex systems initiative of the Future and Emerging Technologies
directorate of the IST Priority, EU Sixth Framework
and by EU grant HPRN-CT-2002-00319,q.


\begin{thebibliography}{}
\bibitem{note2}
The  the Internet datasets  are the one collecetd by University of Oregon Route Views project, NLANR and the protein-protein interaction datasets are one listed in DIP database.
\end{thebibliography}
\end{document}